\documentstyle›12pt!{article}
\title{ONE-DIMENSIONAL DYNAMIC AND\\ FACTORS OF FINITE AUTOMATA}
  \author{Petr K\accent23urka \vspace{8pt}\\
  Dept. of Mathematical Logic and Philosophy of Mathematics,\\
  Faculty of Mathematics and Physics, Charles University\\
  Malostransk\'{e} n\'{a}m\v{e}st\'{\i} 25, 118 00 Praha 1, Czechia\\
  {\small fax: (422) 532 742, email: kurka@cspguk11.bitnet}}
\date{}
\newcommand{\mez}{\vspace{10pt}}
\newcommand{\arleft}›2! {\begin{picture}(100,10)
\put(10,0){$#2$} \put(27,3){\vector(1,0){48}} \put(48,10){$#1$}
\put(83,0){$#2$} \end{picture} }
\newcommand{\arup}›2! {\begin{picture}(100,25)
\put(3,12){$#1$}      \put(94,12){$#2$}
\put(15,2){\vector(0,1){23}} \put(88,2){\vector(0,1){23}}
\end{picture} }
\newcommand{\ardown}›2! {\begin{picture}(100,25)
\put(3,12){$#1$}      \put(94,12){$#2$}
\put(15,25){\vector(0,-1){23}} \put(88,25){\vector(0,-1){23}}
\end{picture}}

\newtheorem{dfn}{Definition}
\newtheorem{lem}{Lemma}
\newtheorem{pro}{Proposition}
\newtheorem{thm}{Theorem}
\newtheorem{cor}{Corollary}

\begin{document}
\maketitle

\begin{abstract}

We argue that simple dynamical systems are factors of finite
automata, regarded as dynamical systems on discontinuum. We show
that any homeomorphism of the real interval is of this class. An
orientation preserving homeomorphism of the circle is a factor of
a finite automaton iff its rotation number is rational.
Any $S$-unimodal system on the real interval, whose kneading
sequence is either periodic odd or preperiodic, is also a factor of a
finite automaton, while $S$-unimodal systems at limits of period
doubling bifurcations are not.

\end{abstract}

\section{Introduction}

The simplest dynamical systems are characterized by the presence
of single fixed points, which attract all other points, the
typical example being the halving of the interval $H(x)=x/2$.
Chaotic systems, on the other hand, keep visiting every region of
the state space in apparently chaotic manner, the simplest
example being the doubling of the circle $D(x) = 2x \mbox{ mod }
1$. Nevertheless, the latter system does not seem to be more
complex, and can  be regarded as a dual, of the former. Indeed if
$x=0.x_{0}x_{1}x_{2}...$ is a binary expansion of $x \in ›0,1!$,
then $H(x) = 0.0x_{0}x_{1}...$, while $D(x) = 0.x_{1}x_{2}...$.
It turns out that both systems are factors of finite automata
regarded as dynamical systems on discontinuum. The halving of the
interval writes zero to the output tape, while the doubling of
the circle reads a digit from the input tape.

These considerations lead to a complexity classification of
dynamical  systems. According to a generalized Alexandrov
Theorem, every dynamical  system on a compact metric space is a
factor of some dynamical system  on discontinuum  (see Balcar and
Simon \cite{kn:Balcar} for a proof). Every type of automata
studied in computer science can be regarded as a class of
dynamical systems on discontinuum. The complexity hierarchy
ranging from finite automata, stack automata, Turing automata,
and cellular automata to neural networks can be transferred to
dynamical systems on compact metric spaces via factorization.
The factors of finite automata are at the bottom of this hierarchy.
We have shown in \cite{kn:Kurka} that they include systems with
finite attractors and hyperbolic systems.

In the present paper we extend and generalize these results.
We use a more general concept of a multitape finite automaton with
variable advance of its tapes. We show that an orientation
preserving homeomorphism of the circle is a factor of a finite
automaton iff its rotation number is rational. We show that any
$S$-unimodal system on real interval with preperiodic or odd
periodic kneading sequence is a factor of a
finite automaton. To obtain these results we modify the kneading
theory using the  two closed intervals on either side of the
critical point instead of the open ones. There are multiple
itineraries in some cases, and we choose as a standard the least
one. This modification substantially simplifies the theory, and
we obtain immediately a homomorphism from itineraries to points.
Our approach is dual to that of Crutchfield and Young
\cite{kn:Crutch}, who investigate the regularity of languages
generated by quadratic systems.

Finally we prove that the $S$-unimodal systems with aperiodic
kneading sequences, which occur at common limits
of period doubling and band merging bifurcations, are not factors
of finite automata.  Thus we leave open the case of even periodic
and arbitrary aperiodic kneading sequences. The negative results
are obtained with the use of a topological property of having chaotic
limits. This is another simplicity criterion for dynamical systems.
The standard Devaney's \cite{kn:Devaney} definition reads that a chaotic
system is one, which is topologically transitive, whose periodic
points are dense, and which is sensitive to initial conditions.
However, Brooks et al.\ \cite{kn:Brooks} have recently proved,
that on infinite metric spaces, the first two conditions imply
the third. In light of this result it is natural to relax the
definition of chaotic system to topological transitivity and
density of periodic points. According to this relaxed definition,
finite dynamical  systems, which are orbits of periodic points,
are chaotic too.  These considerations provide some support for
the feeling that chaotic systems are simple too. We define a
system with chaotic limits as one, whose each point is included
in a set, whose $\omega$-limit set is chaotic. We show that
systems which do not have chaotic limits are not factors of
finite automata.

\section{Systems with chaotic limits}

\begin{dfn} A dynamical system $(X,f)$ is a continuous map
$f: X \rightarrow X$ of a compact metric space $X$ to itself.
A homomorphism $\varphi: (X,f) \rightarrow (Y,g)$ of dynamical
systems is a continuous mapping $\varphi:X \rightarrow Y$ such
that $g \varphi = \varphi f$. We say that $(Y,g)$ is a factor of
$(X,f)$, if $\varphi$ is a factorization (i.e. a surjective map).

\begin{picture}(200,55)
\put(124,45){\arleft{f}{X}}
\put(124,15){\ardown{\varphi}{\varphi}}
\put(124,5) {\arleft{g}{Y}}
\end{picture}
\end{dfn}

\noindent
The $n$-th iteration of $f$ is defined by $f^{0}(x)=x$,
$f^{n+1}(x)=f(f^{n}(x))$. A point $x \in X$ is periodic with
period $n>0$, if $f^{n}(x)=x$ and $f^{i}(x) \neq x$ for $0<i<n$.
A point $x$ is eventually periodic, if there exists $i \geq 0$,
such that $f^{i}(x)$ is periodic. It is preperiodic, if it is
eventually periodic but not periodic. It is aperiodic, if it is
not eventually periodic.

A subset $A \subseteq X$ is invariant, if $f(A) \subseteq  A$.
The $\omega$-limit set of a set $A \subseteq  X$ is
$\omega(A)= \bigcap_{n} \overline{ \bigcup_{m>n} f^{m}(A)}$.
It is easy to see that if $A$ is nonempty, then $\omega(A)$ is a
nonempty, closed, invariant set, so $(\omega(A),f)$ is a
dynamical system. If $\varphi: (X,f) \rightarrow (Y,g)$ is a
homomorphism, then $\varphi(\omega_{f}(A)) =
\omega_{g}(\varphi(A))$.

\begin{dfn}
A dynamical system $(X,f)$ is chaotic, if the set of its periodic
points is dense in $X$ and if for any open nonempty sets $U,V
\subseteq  X$ there exists $k>0$ with $f^{k}(U) \cap V \neq 0$
(topological transitivity). The system $(X,f)$ has chaotic
limits, if for each $x \in X$ there exists $A \subseteq X$, such
that $x \in A$ and $\omega(A)$ is chaotic.
\end{dfn}

\noindent
In particular a finite system is chaotic iff it is an orbit of a
periodic point, and any finite system has chaotic limits. If
$\omega(x)$ is a periodic orbit for each $x \in X$, then $(X,f)$
is a system with chaotic limits. Moreover if $X$ or $\omega(X)$
is chaotic, then $(X,f)$ has chaotic limits. It is easy to see
that a factor of a chaotic dynamical system is chaotic, and a
factor of a system with chaotic limits has chaotic limits.

\section{Finite automata}

Let $A$ be a finite alphabet and $N=\{0,1,...\}$ the set of
non-negative integers. We frequently use alphabet $2=\{0,1\}$.
Denote $A^{N}= \{u = u_{0}u_{1}... º u_{i} \in A \}$ the set of
infinite sequences of letters of $A$,  $A^{*}$ the set of finite
sequences, and $\overline{A^{*}} = A^{*} \cup A^{N}$. Denote
$ºuº$ the length of a sequence $u \in \overline{A^{*}}$,
($0 \leq ºuº \leq \infty)$, and $\wedge$ the word of zero length so,
$u_{i}=\wedge$ for $i>ºuº$. Denote $u_{ºi} = u_{0}...u_{i-1}$ the
initial substring of $u$ of length $i$, and  write $u \sqsubseteq v$,
if $u$ is an initial substring of $v$. If $u \in A^{*}$,
denote $\overline{u} \in A^{N}$ the infinite repetition of $u$,
defined by $\overline{u}_{kºuº+i} = u_{i}$. Define a metric
$\varrho_{A}$ on $\overline{A^{*}}$ by $\varrho_{A}(u,v) =
2^{-n}$, where $n = \inf\{i \geq 0º u_{i} \neq v_{i} \}$.

We define the shift $\sigma: \overline{A^{*}} \rightarrow
\overline{A^{*}}$ by $\sigma(u)_{i}=u_{i+1}$, thus
$\sigma(\wedge) = \wedge$. We use also inverse shifts. To unify
the formalism, suppose that $\bullet, \sharp$ are symbols not
occurring in $A$, and define $\sigma_{a}: A^{N} \rightarrow
A^{N}$ for $a \in A \cup \{\bullet,\sharp\}$ by
\› \begin{array}{ll}
   \sigma_{\sharp}(u_{0}u_{1}u_{2}...) = u_{1}u_{2}u_{3}...\\
   \sigma_{\bullet}(u_{0}u_{1}u_{2}...) = u_{0}u_{1}u_{2}...\\
   \sigma_{a}(u_{0}u_{1}u_{2}...) = au_{0}u_{1}... & \mbox{ for }
     a \in A \end{array} \!
An automaton (finite, stack or Turing) is a finite control
attached to one or more tapes with some rules of access to them.
For formal reasons it is simpler to conceive an automaton as a
finite system of infinite tapes, which interact at their ends.
In general  case these tapes are stacks, so a letter can be either
read or written to them. Two stacks already can be used to
simulate any Turing machine. Finite automata possess only input
or output tapes, which move in one direction only.

\begin{dfn}
Let $T$ be a finite set (of tapes) and for each $t \in T$ let
$A_{t}$ be a finite alphabet. Denote $Q = \prod_{t \in T} A_{t}$,
$X = \prod_{t \in T} A_{t}^{N}$ the state space with a product
metric $\varrho$, and $\pi:X \rightarrow Q$ the projection
defined by $\pi(u)_{t} = u_{t0}$.  An automaton with tapes $T$,
alphabets $A_{t}$ and transition functions
$f_{t}: Q \rightarrow A_{t} \cup \{\bullet,\sharp\}$
is a dynamical system $(X,f)$ defined by
\› f(u)_{t} = \sigma_{f_{t} \pi(u)}(u_{t}) ,\;\; u \in X,\;\; t \in T \!
We say that $t \in T$ is an input tape, if
$\;(\forall q \in Q)(f_{t}(q) \in \{\bullet,\sharp\})$.
We say that $t \in T$ is an output tape, if
$\;(\forall q \in Q)(f_{t}(q) \in A_{t} \cup \{\bullet\})$.
An automaton is a finite automaton, if every tape is either input
tape or output tape.
\end{dfn}

\begin{dfn}
For an automaton $(X,f)$ over $T$, $u \in X$, $n \geq 0$ define
the advance $d_{t}(u,n)$ of $t$ in $n$ steps by $d_{t}(u,0)=0$,
$d_{t}(u,1) = 1$ if $f_{t} \pi (u) = \sharp$, $d_{t}(u,1) = 0$ if
$f_{t} \pi (u) = \bullet$, $d_{t}(u,1) = -1$ if $f_{t} \pi (u)
\in A_{t}$, and $d_{t}(u,n+1) =  d_{t}(u,1) + d_{t}(f(u),n)$  for
$n>0$. Clearly $ºd_{t}(u,n)º \leq n$.
\end{dfn}

\begin{thm} \label{limit}
Any finite automaton has chaotic limits.
\end{thm}
Proof: Let $(X,f)$ be a finite automaton and $w \in X$. Denote
$Q_{0}=\{q \in Qº (\forall m)(\exists n \geq m)(\pi f^{n}(w) = q)
\}$, $T_{in} = \{t \in Tº (\exists q \in Q_{0})(f_{t}(q) = \sharp
) \}$,  $T_{out}=\{t \in Tº(\exists q \in Q_{0})(f_{t}(q) \in
A_{t}) \}$, $T_{\bullet} = T - T_{in} - T_{out}$. Since $(X,f)$
is a finite automaton, $T_{in} \cap T_{out} = 0$. Let $j$ be the
first integer, such that $(\forall n \geq j)(\pi f^{n}(w) \in
Q_{0})$ and denote\\ \hspace*{30pt}
$Y = \{ u \in Xº(\forall i)(\pi f^{i}(u) \in Q_{0}) \;\&\;
(\forall t \in T_{\bullet})(\forall i)(u_{ti} = f^{j}(w)_{ti})
\}$.\\ Then $Y$ is a closed invariant set. Let $p$ be the least
integer, for which there exists $z \in Y$ with $\{\pi f^{i}(z)º i
<p\} = Q_{0}$ and $\pi f^{p}(z) = \pi(z)$. Denote\\
\hspace*{30pt} $W = \{u \in Y º(\forall q \in Q_{0})(\forall k
>0)(º\{i<kpº \pi f^{i}(u) = q \}º \geq k/2 \}) \}$\\
(here $ºAº$ means the number of elements of $A$). We shall prove
in the following lemmas that  $\omega(W \cup \{w\})$ is chaotic.

\begin{lem} \label{limit1}
Let $u \in Y$ and let for all $k$ there exist $n$, $m$, and  $v
\in Y$, such that $\varrho(f^{n}(v),u) \leq 2^{-k}$,
$(\forall t \in T_{out})(d_{t}(v,n) \leq -k)$, and
$(\forall t \in T_{in}) (d_{t}(f^{n}(v),m) \geq k)$. Then
$u \in \omega(W)$.
\end{lem}
Proof: Let $z \in Y$ be such that $\{\pi f^{i}(z) º i <p\} =
Q_{0}$, $\pi(z)=\pi(v)$, and $(\forall i)( \pi f^{p+i}(z) = \pi
f^{i}(z))$. Let $r>n+m+p$ be an integer divisible by $p$. There
exists $s \in Y$ satisfying $\pi(s) = \pi(v)$, and for all $t \in
T_{in}$
\› \begin{array}{lll}
  s_{t,i} = z_{t,i} & \mbox{ for } & 1 \leq i \leq a=d_{t}(z,r)
\\   s_{t,a+i}=v_{t,i} & \mbox{ for } & 1 \leq i \leq
b=d_{t}(v,n+m)\\   s_{t,a+b+i} = z_{t,\ell+i} & \mbox{ for } & 1
\leq i,
 \mbox{ where } \pi f^{n+m}(v)=\pi f^{\ell}(z) \end{array} \!
Then $s \in W$, $f^{r}(s)_{ti} = v_{ti}$ for all $i \leq
d_{t}(v,n+m)$, $t \in T_{in}$, and $\varrho(f^{r+n}(s),f^{n}(v))$
$\leq 2^{-k}$, so $\varrho(f^{r+n}(s),u) \leq 2^{-k}$, and
therefore $u \in \omega(W)$. $\Box$

\begin{lem}
$\omega(W \cup\{w\}) = \omega(W)$.
\end{lem}
Proof: It suffices to show $\omega(w) \subseteq  \omega(W)$. Let
$u \in \omega(w)$ and $k>0$. We have $f^{j}(w) \in Y$ and there
exist $n,m$, such that $d_{t}(f^{j}(w),n) \leq -k$ for all  $t
\in T_{out}$, $\varrho(f^{j+n}(w),u) \leq 2^{-k}$, and
$d_{t}(f^{j+n}(w),m) \geq k$. By Lemma \ref{limit1},
$u \in \omega(W)$. $\Box$.

\begin{lem}
Periodic points are dense in $\omega(W)$.
\end{lem}
Proof: Let $U \subseteq X$ be an open set, and $U \cap \omega(W)
\neq 0$. There exists $u \in U$ and $k$ such that $u \in C_{k}(u)
\cap \omega(W) \subseteq U \cap \omega(W)$, where $C_{k}(u) =  \{
v \in Xº u_{ºk}=v_{ºk} \}$ is a cylinder around $u$. There exist
$m$ and $v \in W$ such that $\varrho(f^{m}(v),u) \leq 2^{-k}$,
and $ºd_{t}(v,m)º \geq k$ for all $t \in T_{in} \cup T_{out}$.
There exists $r>m+k$ and $s \in Y$ such that $\pi(s) = \pi(v)$, and
\› \begin{array}{ll}
  s_{t,i} = v_{t,i} & \mbox{ for } t \in T_{in}, \;\;
            1 \leq i \leq d_{t}(v,m)+k \\
  \{\pi f^{i}(s)º i<r\} = Q_{0} \\
  \pi f^{r}(s) = \pi(v),\\
  s_{t,a+i} = s_{t,i} & \mbox{ for } t \in T_{in},\;\; 1 \leq i, \;\;
            a = d_{t}(s,r) \\
  s_{t,i} = f^{r}(s)_{t,i} & \mbox{ for } t \in T_{out}
\end{array} \!
Then $s \in Y$, $f^{r}(s)=s$, $ºd_{t}(s,r)º \geq k$ for
$t \in T_{in} \cup T_{out}$, and $\varrho(f^{m}(s),u) \leq
2^{-k}$. By Lemma \ref{limit1}, $s \in \omega(W)$. $\Box$

\begin{lem}
$\omega(W)$ is topologically transitive.
\end{lem}
Proof: Let $U,U' \subseteq  X$ be open sets, $u \in C_{k}(u) \cap
\omega(W) \subseteq U \cap \omega(W)$, $u' \in C_{k}(u') \cap
\omega(W) \subseteq U' \cap \omega(W)$. There exist $m,m'$ and
$v,v' \in W$ with $\varrho(f^{m}(v),u) \leq 2^{-k}$,
$\varrho(f^{m'}(v'),u') \leq 2^{-k}$, $ºd_{t}(v,m)º \geq k$,
$ºd_{t}(v',m')º \geq k$ for all $t \in T_{in} \cup T_{out}$.
There exist $r>m+k$, $r'>r+m'+k$ and $s \in Y$ such that
\› \begin{array}{ll}
  s_{t,i} = v_{t,i} & \mbox{ for } t \in T_{in},\;\;
        1 \leq i \leq d_{t}(v,m)+k \\
  \pi f^{r}(s) = \pi(v'),\\
  s_{t,a+i} = v'_{t,i} & \mbox{ for } t \in T_{in},\;\;
    1 \leq i \leq d_{t}(v',m')+k,\;\; a = d_{t}(s,r) \\
  \{\pi f^{i}(s)º i<r'\} = Q_{0} \\
  \pi f^{r'}(s) = \pi(v),\\
  s_{t,b+i} = s_{t,i} & \mbox{ for } t \in T_{in},\;\; 1 \leq i, \;\;
  b = d_{t}(s,r') \\
  s_{t,i} = f^{r'}(s)_{t,i} & \mbox{ for } t \in T_{out}
  \end{array} \!
Then $f^{m}(s) \in U$, $f^{r+m'}(s) \in V$, and by Lemma
\ref{limit1}, $s \in \omega(W)$. $\Box$

\section{Homeomorphisms}

\begin{lem} \label{increase1}
Let $g : I \rightarrow  I$ be an increasing continuous function
on a compact real interval $I$, whose only fixed point is one of
the endpoints of $I$. Then $(I,g)$ is a factor of
$\;(2^{N},\sigma_{0})$.
\end{lem}
Proof: Let $I=›a,b!$, $g(a)=a$, $g(b)<b$ (the dual case $g(b)=b$
is similar). For $k \geq 0$ denote $I_{k}=›g^{k+1}(b),g^{k}(b)!$,
$X_{k}=\{u \in 2^{N}º u_{ºk+1}=0^{k}1\}$. For $u \in X_{0}$
define $\varphi(u) = g(b)+(b-g(b)) \sum_{i=1}^{\infty} u_{i}2^{-i}$.
For $u \in X_{k}$, $k>0$ define $\varphi(u) = g^{k} \varphi
\sigma^{k}(u)$, $\varphi(\overline{0}) = a$. Then
$\varphi : (2^{N},\sigma_{0}) \rightarrow (I,g)$ is a
factorization.  $\Box$

\begin{lem} \label{increase2}
Let $(X,f)$ be a two tape finite automaton defined by $f(u,v)$ =
$(\sigma_{\sharp}(u), \sigma_{\max(u_{0},v_{0})}(v))$. Let $g : I
\rightarrow  I$ be an increasing continuous function on a compact
real interval $I$, whose only fixed points are the two endpoints
of $I$. Then $(I,g)$ is a factor of $(X,f)$.
\end{lem}
Proof: Let $I=›a,b!$ and pick some $c \in (a,b)$. Suppose
$f(c)>c$ (the dual case $f(c)<c$ is similar). For $k \in Z$
denote
$I_{k}=›g^{k}(c),g^{k+1}(c)!$, There exist closed sets
\› \begin{array}{llll}
   X_{-\infty} & = & \{(u,v) \in Xº v_{0}=0 \;\;\&\;\;
         (\forall i \geq 0)(u_{i}=0) \}\\
   X_{k} & = & \{(u,v) \in Xº v_{0}=0 \;\;\&\;\;
         \min\{i \geq 0º u_{i}=1\} = -k \} , &   k \leq 0 \\
   X_{k} & = & \{(u,v) \in Xº \min\{i \geq 0º v_{i}=0\} = k \} , &
          k > 0 \\
   X_{\infty} & = & \{(u,v) \in Xº (\forall i \geq 0)(v_{i}=1) \}
\end{array} \!
For $k \in Z$ define a projection $\pi_{k}: X_{k} \rightarrow
2^{N}$ by $\pi_{k}(u,v) = \sigma^{k}_{\sharp}(v)$ if $k \geq 0$,
and $\pi_{k}(u,v) = \sigma^{-k}_{0}(v)$ if $k \leq 0$. Then
$\pi_{k+1}f(u,v)=\pi_{k}(u,v)$ for $(u,v) \in X_{k}$. Let
$\psi : 2^{N} \rightarrow I_{0}$ be a factorization. Define
$\varphi(u,v) = g^{k} \psi \pi_{k}(u,v)$ for $(u,v)
\in X_{k}$, $\varphi(u,v)=a$ for $(u,v) \in X_{-\infty}$, and
$\varphi(u,v) = b$ for $(u,v) \in X_{\infty}$. Then $\varphi:
(X,f) \rightarrow (I,g)$ is a factorization. $\Box$

\begin{pro} \label{increase}
Let $g: I \rightarrow I$ be a non-decreasing continuous map on a
compact real interval $I$. Then $(I,g)$ is a factor of a finite
automaton.
\end{pro}
Proof: Define a finite automaton $(X,h)$, where $X = 2^{N} \times
2^{N} \times 2^{N} \times 2^{N}$, and  $h(y,u,v,w)$ =
$(\sigma_{0}(y),f(u,v),w)$, where $f$ is the finite automaton
from Lemma \ref{increase2}. We shall prove that $(I,g)$ is a
factor of $(X,h)$. Denote $F=\{x \in Iº g(x)=x\}$ the set of
fixed points. $F$ is closed, therefore $I-F = \cup_{k \in K}
A_{k}$ is a finite or countable union of open intervals
$A_{k}=(a_{k},b_{k})$. Define $\theta :2^{N} \rightarrow I$ by
$\theta(u) = a + (b-a) \sum_{i=0}^{\infty} u_{i}2^{-i+1}$,
where $I=›a,b!$. Each $\theta^{-1}(A_{k})$ contains a cylinder
$C_{k} \subseteq  2^{N}$. Let $\prec$ be a lexicographic ordering
on $2^{N}$. Denote
$K_{0} = \{ k \in K º g(a_{k})=a_{k},\; g(b_{k})=b_{k} \}$.
If $k \not\in K_{0}$, let $\psi_{k} : (2^{N}, \sigma_{0}) \rightarrow
(\overline{A_{k}},g)$ be a factorization from Lemma \ref{increase1}.
If $k \in K_{0}$, let $\psi_{k} : (2^{N} \times 2^{N},f)
\rightarrow (\overline{A_{k}},g)$ be a factorization from Lemma
\ref{increase2}. Define $\varphi : (X,h) \rightarrow (I,g)$ by
\› \begin{array}{lllll}
\varphi(y,u,v,w) = \theta(w) & \mbox{ if } & \theta(w) \in F \\
\varphi(y,u,v,w) = a_{k} & \mbox{ if } & \theta(w) \in A_{k}, &
\mbox{and } (\forall x \in C_{k})(w \prec x) \\
\varphi(y,u,v,w) = b_{k} & \mbox{ if } & \theta(w) \in A_{k}, &
\mbox{and } (\forall x \in C_{k})(x \prec w) \\
\varphi(y,u,v,w) = \psi_{k}(y) & \mbox{ if } & w \in C_{k}, &
\mbox{and } k \not\in K_{0} \\
\varphi(y,u,v,w) = \psi_{k}(u,v) & \mbox{ if } & w \in C_{k}, &
\mbox{and } k \in K_{0} & \Box
\end{array} \!

\begin{thm} \label{realhom}
Any homeomorphism of a real interval into itself is a factor of a
finite automaton.
\end{thm}
Proof: If $g: I \rightarrow I$ is increasing, then it is a factor
of finite automaton by Proposition \ref{increase}. If it is
decreasing, then $g^{2}:I \rightarrow I$ is increasing and
therefore there is a factorization $\psi : (X,f) \rightarrow
(I,g^{2})$, and $(X,f)$ is a finite automaton. We add to this
automaton an output tape with alphabet $2$ and define $h(u,0v)$ =
$(u,10v)$, $h(u,1v)$ = $(f(u),01v)$ ($u \in X$, $v \in 2^{*}$).
Then $h: X \times 2 \rightarrow X \times 2$ is a finite
automaton. Define $\varphi(u,0v) = \psi(u)$, and $\varphi(u,1v) =
g \psi(u)$. Then $\varphi : (X \times 2 , h) \rightarrow (I,g)$
is a factorization. Indeed $\varphi h(u,0v) = \varphi(u,10v) = g
\psi(u) = g \varphi(u,0v)$, and $\varphi h(u,1v) =
\varphi(f(u),01v) = \psi f(u) = g^{2}\psi(u) = g\varphi(u,1v)$.
$\Box$ \mez

We turn now to the orientation preserving homeomorphisms of the
circle, which we regard as the unit circle in the complex plane.
For $x \in R_{1}$ denote $\theta(x) = \exp(2\pi i x)$.  An
increasing map $G: R_{1} \rightarrow R_{1}$ is a lift of an
orientation preserving homeomorphism $(S_{1},g)$, if $\theta :
(R_{1},G) \rightarrow (S_{1},g)$ is a factorization. The rotation
number $\varrho(g)$ of $(S_{1},g)$ is defined as the fractional
part of $\varrho_{0}(G) = \lim_{n \rightarrow \infty}
ºG^{n}(y)º/n$, where $G$ is any lift of $g$, and $y \in R_{1}$
($\varrho(g)$ does not depend on the choice of either $G$ or
$y$). Then an orientation preserving homeomorphism of the circle
$(S_{1},g)$ has a periodic point iff $\varrho(g)$ is rational
(see Devaney \cite{kn:Devaney} for a proof).

\begin{thm}
An orientation preserving homeomorphism $(S_{1},g)$ of the circle
is a factor of a finite automaton iff $\varrho(g)$ is rational.
\end{thm}
Proof: If $(S_{1},g)$ is a factor of a finite automaton, then it
has chaotic limits by Theorem \ref{limit}, and therefore also
periodic points. Thus its rotation number cannot be irrational.
If $\varrho(g)$ is rational, then $\varrho(g^{n}) = 0$ for some
$n$, and $g^{n}$ has a fixed point. We can suppose that 1 is a
fixed point of $g$, and choose a lift $G$, with fixed points $0$
and $1$. Then $(S_{1},g^{n})$ is a factor of $(›0,1!,G)$, and
this is a factor of a finite automaton by Theorem \ref{realhom}.
We modify the finite automaton obtained by letting it act only
every $n$-th step similarly as in the proof of Theorem
\ref{realhom}. Then $(S_{1},g)$ is a factor of this modified
automaton. $\Box$

\section{Unimodal systems}

\begin{dfn}
A dynamical system $(I,g)$ defined on real interval $I=›a,b!$ is
unimodal, if $g(a)=g(b)=a$, and if there exists a critical point
$c \in (a,b)$, such that $g$ is increasing in $›a,c!$, and
decreasing in $›c,b!$.  \end{dfn}

\begin{pro}
Let $g$ be unimodal on $›a,b!$ with critical point $c$. Denote
$I_{0}=›a,c!$, $I_{1}=›c,b!$, and $I_{u} = \{x \in I º 0 \leq i <
ºuº \Rightarrow f^{i}(x) \in I_{u_{i}} \}$ for $u \in
\overline{2^{*}}$. Then $I_{u}$ are closed intervals (possibly
empty or singletons).
\end{pro}
Proof:  $I_{au} = I_{a} \cap f^{-1}(I_{u})$ for $a \in 2$ and $u
\in \overline{2^{*}}$. $\Box$

\begin{dfn}
For $u \in \overline{2^{*}}$, $n<ºuº$ denote $\tau_{n}(u) =
\sum_{i=0}^{n} u_{i} \mbox{ mod } 2$. We say that $u \in 2^{*}$
is even (odd), if $\tau_{ºuº-1}(u)$ is zero (one). An infinite
periodic sequence $u \in 2^{\infty}$ is even (odd), if the
shortest sequence $v \in 2^{*}$ with $u = \overline{v}$ is even
(odd). For $u,v \in \overline{2^{*}}$ define
\› \begin{array}{lll}
  u \prec v  & \mbox{ iff } & (\exists k < \min(ºuº,ºvº))
       (u_{ºk}=v_{ºk} \;\;\&\;\; \tau_{k}(u) < \tau_{k}(v))\\
  u \preceq v & \mbox{ iff } & u \prec v \;\mbox{ or }\;
    u \sqsubseteq v  \end{array} \!
\end{dfn}

\begin{pro} \label{interval}
Let $u,v \in \overline{2^{*}}$. Then
\› \begin{array}{llllll}
  &   u \prec v & \Rightarrow & (\forall x \in I_{u})
    (\forall y \in I_{v})(x \leq y) &
    \stackrel{def}{\Leftrightarrow} & I_{u} < I_{v}\\
  I_{u} \neq 0 \;\;\;\& & u \preceq v  & \Rightarrow &
     (\forall y \in I_{v})(\exists x \in I_{u})(x \leq y)
     & \stackrel{def}{\Leftrightarrow} & I_{u} \leq I_{v}
\end{array} \!
\end{pro}
Proof: Let $u \prec v$, $k< ºuº$, $k < ºvº$, $u_{ºk}=v_{ºk}$,
$\tau_{k}(u) < \tau_{k}(v)$, $x \in I_{u}$, $y \in I_{v}$. We
prove $x \leq y$ by induction on $k$. If $k=0$, then $u_{0}=0$,
$v_{0}=1$, so $I_{u} \subseteq I_{0} < I_{1} \supseteq I_{v}$,
and $I_{u} < I_{v}$. If $k>0$, we apply the Proposition to
$\sigma(u)$ and $\sigma(v)$. If $u_{0}=v_{0}=0$, then $\sigma(u)
\prec \sigma(v)$, so $f(x) \leq f(y)$, and $x \leq y$ since $x,y
\in I_{0}$. If $u_{0}=v_{0}=1$, then $\sigma(v) \prec \sigma(u)$,
so $f(y) \leq f(x)$, and $x \leq y$ since $x,y \in I_{1}$.
Suppose $u \preceq v$, and $y \in I_{v}$. Since $I_{u} \neq 0$,
we have some $x \in I_{u}$. If $u \prec v$, then $x \leq y$ by
the above proof. If $u \sqsubseteq v$, then $y \in I_{u}$, and $y
\leq y$. $\Box$

\begin{dfn}
The itinerary ${\cal I}(x) \in 2^{N}$ of $x \in I$ is defined by
\› \begin{array}{lll}
  g^{i}(x) < c & \Rightarrow & {\cal I}(x)_{i}=0\\
  g^{i}(x) > c & \Rightarrow & {\cal I}(x)_{i}=1\\
  g^{i}(x) = c & \Rightarrow & \tau_{i}({\cal I}(x))=0
\end{array} \!
The kneading sequence of $g$ is ${\cal K}(g) = {\cal I}(g(c))$.
\end{dfn}

\begin{pro} \label{order}
\› \begin{array}{rcl}
  x \in I_{u} & \Rightarrow & {\cal I}(x) \preceq u\\
  {\cal I}(x) \prec {\cal I}(y) & \Rightarrow  & x \leq y \\
  x < y & \Rightarrow & {\cal I}(x) \preceq {\cal I}(y)
 \end{array} \! \end{pro}
Proof: The first formula follows from the definition. If ${\cal
I}(x) \prec {\cal I}(y)$, then $I_{{\cal I}(x)} < I_{{\cal
I}(y)}$. Since $x \in I_{{\cal I}(x)}$, $y \in I_{{\cal I}(y)}$,
we get $x \leq y $ by Proposition \ref{interval}. If $x<y$, then
${\cal I}(y) \not\prec {\cal I}(x)$, so ${\cal I}(x) \prec {\cal
I}(y)$. $\Box$

\begin{dfn} For $u,v \in \overline{2^{*}}$ define $u \ll v$ iff
$\;(\forall i>0)(\sigma^{i}(u) \preceq v)$, and $u \ll^{*} v$ if
$u \ll v$ and $u \preceq v$. A sequence $w \in \overline{2^{*}}$
is maximal, if $\;w \ll w$.
\end{dfn}

\begin{thm} \label{knead1}
If $(I,g)$ is unimodal, then
$(\forall x \in I)({\cal I}(x) \ll {\cal K}(g))$, and ${\cal
K}(g)$ is maximal.
\end{thm}
Proof: Let $x \in I$ and $k > 0$. Denote $u = {\cal I}(x)$ and
suppose $\sigma^{k}(u) \succ {\cal K}(g)$. We have $g^{k}(x) \leq
g(c)$. If $g^{k}(x) = g(c)$, then $g^{k-1}(x) = c$, and
$\tau_{k-1}({\cal I}(x)) = 0$, so $\sigma^{k}(u) = {\cal
I}(g^{k}(x)) = {\cal K}(g)$, which is a contradiction. Thus
$g^{k}(x) < g(c)$, so $g^{k-1}(x) \neq c$, and by Proposition
\ref{order}, ${\cal I}(g^{k}(x)) \preceq {\cal K}(g)$. Let
$i \geq k$ be the least number with $g^{i}(x)=c$ (such a number
exists, since otherwise ${\cal I}(g^{k}(x)) = \sigma^{k}(u)$).
There exists $y \in I$ with ${\cal I}(y)=0u_{k}...u_{i}...$ and
$f^{j}(y) \neq c$ for $0 \leq j \leq i-k+1$. Then $f(y) <f(c)$
and ${\cal I}(f(y)) = u_{k}...u_{i}... \succ {\cal K}(g)$, which
is a contradiction. $\Box$
\pagebreak

\begin{thm} \label{knead2}
If $(I,g)$ is unimodal, $u \in \overline{2^{*}}$, and $u \ll
{\cal K}(g)$, then $I_{u} \neq 0$.
\end{thm}
Proof: We proceed by induction on $n=ºuº$. If $n \leq 1$, then
$I_{u} \neq 0$. Suppose that the Theorem holds for
$n-1$. Then the assumption holds for $\sigma(u)$, and therefore
$I_{\sigma(u)} \neq 0$. Since $\sigma(u) \preceq {\cal K}(g)$,
and $g(c) \in I_{{\cal K}(g)}$, there exists by Proposition
\ref{interval} some $y \in I_{\sigma(u)}$ with $y \leq g(c)$.
There exists $x \in I_{u_{0}}$ with $g(x) = y$, and therefore $x
\in I_{u} \neq 0$. Thus we have proved the Theorem for finite
$ºuº$. If $ºuº = \infty$, then  $I_{u} = \cap \{I_{v}º
v \sqsubseteq u, \; v \in 2^{*}\}$. Since $v \sqsubseteq u$
implies $v \ll {\cal K}(g)$, and therefore $I_{v} \neq 0$, we get
$I_{u} \neq 0$ by compactness. $\Box$

\begin{dfn}
A unimodal system $(I,g)$ is $S$-unimodal, if it has continuous
third derivation, $g''(c)<0$, and $Sg(x)<0$ for all $x \neq c$,
where $Sg(x) = \frac{g'''(x)}{g'(x)} -
\frac{3}{2}(\frac{g''(x)}{g'(x)})^{2}$ is the Schwarzian
derivative.
\end{dfn}

\begin{thm} \label{guck}
Suppose that $g$ is $S$-unimodal. If ${\cal K}(g)$ is not
periodic, then $I_{u}$ has empty interior for any $u \in 2^{N}$.
If ${\cal K}(g)$ is periodic with period $n$, and $u \in 2^{N}$,
then $I_{u}$ has nonempty interior iff $u \ll {\cal K}(g)$ and
there exists $k \geq 0$ with $\sigma^{k}(u) = {\cal K}(g)$.
\end{thm}
See Guckenheimer \cite{kn:Gucken} or Collet and Eckmann
\cite{kn:Collet} for a proof.

\begin{dfn}
Let $w \in 2^{N}$ be maximal and define $\Sigma_{w} = \{u \in
2^{N}º u \ll w\}$, $\Sigma_{w}^{*} = \{u \in 2^{*}º u \ll w \}$.
Then $\Sigma_{w}$ is a subshift, i.e. a closed
$\sigma$-invariant subset.
\end{dfn}

\begin{thm} \label{nonper}
Let $(I,g)$ be S-unimodal system such that ${\cal K}(g)$ is not
periodic. Then $(I,g)$ is a factor of $(\Sigma_{{\cal
K}(g)},\sigma)$.
\end{thm}
Proof: By Theorems \ref{knead2} and \ref{guck}, $I_{u}$ is a
singleton for each $u \in \Sigma_{{\cal K}(g)}$, so we define
$\varphi: \Sigma_{{\cal K}(g)} \rightarrow I$ by $\varphi(u) \in
I_{u}$. By Theorem \ref{knead1}, $\varphi$ is surjective. Let us
prove that it is continuous. If $\varphi(u) \in U$, and $U$ is
open, then there is $v \sqsubseteq u$ with $v \in 2^{*}$ and
$I_{v} \subseteq  U$. It follows that for all $w \in 2^{N}$ with
$v \sqsubseteq w$ we have $\varphi(w) \in U$. $\Box$
\pagebreak

\section{Eventually periodic kneading sequences}

\begin{thm} \label{presigma}
If $w \in 2^{N}$ is preperiodic and maximal, then
$(\Sigma_{w},\sigma)$ is a factor of a finite automaton.
\end{thm}
Proof: We can write $w=w_{0}...w_{m-1}\overline{w_{m}...w_{m+n-1}}$,
where $w_{m-1} \neq w_{m+n-1}$, and $w_{m}...w_{m+n-1}$ is even.
Then $\sigma^{i}(w) \prec w$ for all $i>0$, and since the set
$\{\sigma^{i}(w) º i>0\}$ is finite, there exists $p$ such that
$\sigma^{i}(w)_{ºp} \prec w_{ºp}$ for all $i>0$. Let $k$ be an
integer with $nk>p$ and put $r=m+nk$, $s=m+(n+1)k$.  Assume
$w_{ºr}u \ll^{*} w$ and let us prove $\sigma^{i}(w_{ºs}u) \preceq w$
for all $i \geq 0$. For $i=0$ we distinguish two cases. If $w_{ºm}$
is even then $w_{ºr}$ is even, and $u \preceq \sigma^{r}(w) =
\sigma^{s}(w)$ by the assumption, so $w_{ºs}u \preceq w$.
If $w_{ºm}$ is odd then $u \succeq
\sigma^{r}(w) =\sigma^{s}(w)$, and again $w_{ºs}u \preceq w$. For
$0< i <s-p$ we have $\sigma^{i}(w_{ºs}u) \prec w$ by
$\sigma^{i}(w)_{ºp} \prec w_{ºp}$. For $i \geq s-p$ we have $i >
n$ and $\sigma^{i}(w_{ºs}u) = \sigma^{i-n}(w_{ºr}u) \preceq w$.
Conversely, if $w_{ºs}u \ll^{*} w$, then $w_{ºr}u \ll^{*} w$.
By induction, $(\forall u \in 2^{N})(w_{ºr}u \ll^{*} w
\Leftrightarrow w_{ºr+jn}u \ll^{*} w)$ for all $j>0$.
We construct a finite automaton with input alphabet $A_{0}=2$, and
output alphabets $A_{1}=2$, $A_{2}=\{0,...,s-1\}$. Put $f_{0}(q)
= \sharp$, $f_{1}(q) = q_{0}$ if $w_{ºq_{2}}q_{0} \preceq w$,
$f_{1}(q) = 1-q_{0}$ otherwise. Then $w_{ºq_{2}}f_{1}(q) \preceq
w$. If $w_{ºq_{2}}f_{1}(q) = w_{ºq_{2}+1}$, then put $f_{2}(q) =
q_{2}+1$ if $q_{2}+1 < s$, and $f_{2}(q) = r$ if $q_{2}+1 = s$.
If $w_{ºq_{2}}f_{1}(q) \prec w_{ºq_{2}+1}$, then put
$f_{2}(q) = \max\{k<q_{2}+1º \sigma^{q_{2}+1-k}(w_{ºq_{2}}f_{1}(q))
= w_{ºk} \}$. Define $\varphi: X \rightarrow \Sigma_{w}$ by
$\varphi(u)_{i} = (\pi f^{i}(u))_{1}$. Then $\varphi$ is
continuous and $\varphi f = \sigma \varphi$. Let $v \in
\Sigma_{w}$, and put $q_{1}=v_{0}$, $u_{0i} = v_{i+1}$,
$q_{2}=0$. Then $\varphi(u) = v$, so $\varphi$ is surjective.
$\Box$

\begin{cor}
If $(I,g)$ is $S$-unimodal and ${\cal K}(g)$ is preperiodic, then
$(I,g)$ is a factor of a finite automaton.
\end{cor}
Proof: Theorems \ref{nonper} and \ref{presigma}.

\begin{pro} \label{sigma}
If $w \in 2^{N}$ is odd periodic and maximal, then $(\Sigma_{w},
\sigma)$ is a factor of a finite automaton. Moreover, if
$u \in \Sigma_{w}$, and $\sigma^{i}(u) = w$ for some $i \geq 0$,
then $u$ is an isolated point of $\Sigma_{w}$.
\end{pro}
Proof: Let $n$ be the period of $w$. Then $(\forall u \in 2^{N})
(w_{ºn}u \ll^{*} w \Leftrightarrow u=w)$. Indeed if $w_{ºn}u \ll^{*}
w$, then $\sigma^{n}(w_{ºn}u) = u \preceq w$ by definition, and
$u \succeq \sigma^{n}(w) =w$, since $w_{ºn}$ is odd, thus $u=w$.
Thus $w$ is an isolated point of $\Sigma_{w}$, and if $\sigma^{i}(u)
= w$, then $u$ is isolated too. Construct a finite automaton with
input alphabet $A_{0}=2$, and output alphabets $A_{1}=2$,
$A_{2}=\{0,...,2n-1\}$. Put $f_{0}(q) = \sharp$, and suppose
$q_{2}<n$. If $w_{ºq_{2}}q_{0} \prec w$, then put $f_{1}(q)=q_{0}$,
$f_{2}(q) = \max\{k<q_{2}+1º \sigma^{q_{2}+1-k}(w_{q_{2}}f_{1}(q))
= w_{ºk} \}$. If $w_{ºq_{2}}q_{0} = w_{ºq_{2}+1}$, then put
$f_{1}(q)=q_{0}$, $f_{2}(q) = q_{2}+1$. If $w_{ºq_{2}}q_{0} \succ w$,
then put $f_{1}(q)=1-q_{0}$, $f_{2}(q) = q_{2}+1$. If $n \leq q_{2} <2n-1$,
then put $f_{1}(q)=w_{q_{2}+1}$,  $f_{2}(q)=q_{2}+1$. If $q_{2}=2n-1$,
then put $f_{1}(q)=w_{q_{2}+1}$,  $f_{2}(q)=n$. Define $\varphi: X
\rightarrow \Sigma_{w}$ by $\varphi(u)_{i} = (\pi f^{i}(u))_{1}$.
Then $\varphi : (X,f) \rightarrow (\Sigma_{w},\sigma)$ is a factorization.
$\Box$

\begin{thm}
If $(I,g)$ is $S$-unimodal and ${\cal K}(g)$ is periodic and odd,
then $(I,g)$ is a factor of a finite automaton.
\end{thm}
Proof: Let $n$ be the period of ${\cal K}(g)$. By Proposition
\ref{sigma}, there is a factorization $\varphi' : (X',f')
\rightarrow (\Sigma_{{\cal K}(g)},\sigma)$.
By Theorem  \ref{realhom} there is a factorization
$\varphi'':(X'',f'') \rightarrow (I_{{\cal K}(g)},g^{n})$.
We construct a finite automaton $(X,f)$, where $X = X' \times X''$,
and $f(u',u'') = (f'(u'),f''(u''))$ if $q'_{2}=0$, or $q'_{2}=n$,
$f(u',u'') = (f'(u'),u'')$ otherwise.
Thus if $\sigma \varphi'(u') = {\cal K}(g)$, then either
$q'_{2}=0$, or $q'_{2}=n$, so $f(u',u'') = (f'(u'),f''(u''))$.
Define $\varphi: X \rightarrow I$ by
\› \begin{array}{lll}
\varphi(u',u'') \in I_{\varphi'(u')} & \mbox{ if } &
   (\forall i)(\sigma^{i}\varphi'(u') \neq {\cal K}(g)) \\
\varphi(u',u'') = g_{\varphi'(u')_{0}}g^{n}\varphi''(u'')  &
\mbox{ if } &      \sigma \varphi'(u') = {\cal K}(g) \\
\varphi(u',u'') = g_{\varphi'(u')_{0}} \varphi(f'(u'),u'')  &
\mbox{ if } & \varphi(f'(u'),u'') \mbox{ has been defined}
\end{array} \!
(Here $g_{i}: ›a,g(c)! \rightarrow I_{i}$, are the two inverses of $g$.)
Then $\varphi(u',u'') \in I_{\varphi'(u')}$. Since $\varphi''$ is
continuous, and since by Proposition \ref{sigma} $\varphi'(u')$
is an isolated point when $\sigma^{i}\varphi'(u') = {\cal K}(g)$,
$\varphi$ is continuous. By Theorems \ref{knead1} and \ref{guck},
$\varphi$ is a surjection.  If $\sigma\varphi'(u')={\cal K}(g)$,
then  $\varphi f(u',u'')$ = $\varphi(f'(u'),f''(u''))$ =
$g_{\varphi'f'(u')_{0}} \varphi((f')^{2}(u'),f''(u''))$ =
$g_{\varphi'(u')_{1}} ... g_{\varphi'(u')_{n-1}}
\varphi((f')^{n}(u'),f''(u''))$ = \\
$g_{\varphi'(u')_{1}} ... g_{\varphi'(u')_{n}} g^{n} \varphi''
f''(u'')$ = $\varphi'' f''(u'')$ = $g^{n} \varphi''(u'')$ =  $g
\varphi(u',u'')$. If $\varphi'(u')$ is eventually periodic, and
if $\sigma\varphi'(u') \neq {\cal K}(g)$, then $\varphi f(u',u'')
= \varphi(f'(u'),u'') = g \varphi(u',u'')$.
Thus $\varphi$ is a factorization. $\Box$

\section{Aperiodic kneading sequences}

For $s \in 2^{*}$, denote $\hat{s}$ the sequence obtained from
$s$ by changing the last bit. The double of $s$ is ${\cal D}(s) =
s\hat{s}$, The doubling operator can be iterated, and since $s
\sqsubseteq {\cal D}(s)$, ${\cal D}^{\infty}(s) =
\lim_{i \rightarrow \infty} {\cal D}^{i}(s)$ is well defined.
It can be easily proved that if $s \in 2^{*}$ is an odd, aperiodic
maximal sequence, then $\overline{\hat{s}} \prec \overline{s} \prec
\overline{{\cal D}(s)} \prec {\cal D}^{\infty}(s)$, and
${\cal D}^{\infty}(s) \in 2^{N}$ is a maximal aperiodic sequence.

\begin{lem} \label{double}
Let $s \in 2^{*}$ be an odd, aperiodic, maximal sequence, $j>2$,
and $U_{j} = \{u \in \Sigma_{{\cal D}^{\infty}(s)} º
(\exists i \leq 2^{j+1}ºsº)({\cal D}^{j}(s) \sqsubseteq
\sigma^{i}(u) \;\mbox{ or } \; \widehat{{\cal D}^{j}(s)}
{\cal D}(s) \sqsubseteq \sigma^{i}(u) \}$. Then $U_{j}$ is a
$\sigma$-invariant set, which is closed and open in
$\Sigma_{{\cal D}^{\infty}(s)}$, and does not contain periodic
points with periods less than $2^{j}ºsº$.
\end{lem}
Proof: Clearly $U_{j}$ is closed and open in $\Sigma_{{\cal
D}^{\infty}(s)}$. To prove it is invariant denote $w = {\cal
D}^{j-2}(s)$ and suppose $u \in U_{j}$. Then we can write
$\sigma^{i}(u)=u_{0}u_{1}...$, where $ºu_{i}º = 2^{j-2}ºsº$.
We use repeatedly the fact $u \ll {\cal D}^{\infty}(s)$ and
prove that if $u_{0}u_{1}u_{2}u_{3} = w\hat{w}ww$, then either
$u_{4}u_{5}u_{6}u_{7}=w\hat{w}ww$ or $u_{4}...u_{11} =
w\hat{w}w\hat{w}w\hat{w}ww$. Since $u_{0}u_{1}u_{2}u_{3}u_{4} \preceq
w\hat{w}www$, and $w\hat{w}ww$ is odd, $u_{4} \succeq w$. Since
$u_{4} \preceq w$, we get $u_{4}=w$. Since $u_{4}u_{5} \preceq
w\hat{w}$, $u_{5} \succeq \hat{w}$. Since $u_{0}...u_{5} \preceq
w\hat{w}www\hat{w}$, and $w\hat{w}www$ is even, $u_{5} \preceq
\hat{w}$, so $u_{5}=\hat{w}$. Since $u_{4}u_{5}u_{6} \preceq
w\hat{w}w$, $u_{6}=w$. Since $u_{6}u_{7} \preceq w\hat{w}$,
$u_{7} \succeq \hat{w}$, so either $u_{7}=w$ or $u_{7}=\hat{w}$.
In the former case we are done. Suppose therefore $u_{7}=\hat{w}$.
Since $u_{0}...u_{8} \preceq w\hat{w}www\hat{w}w\hat{w}w$,
$u_{8}=w$. Since $u_{8}u_{9} \preceq w\hat{w}$, $u_{9} \succeq
\hat{w}$. Since $u_{0}...u_{9} \preceq
w\hat{w}www\hat{w}w\hat{w}w\hat{w}$, $u_{9} \preceq \hat{w}$, so
$u_{9}=\hat{w}$. Since $u_{8}u_{9}u_{10} \preceq w\hat{w}w$,
$u_{10}=w$. Since $u_{0}...u_{11} \preceq
w\hat{w}www\hat{w}w\hat{w}w\hat{w}ww$, $u_{11}=w$, so
$u_{8}u_{9}u_{10}u_{11} = w\hat{w}ww$. If there is a periodic
point $u \in U_{j}$ with period $k < n = ºsº2^{j}$, then
$u_{2n-1}$ = $u_{2n-1-k}$ = $u_{n-1-k}$ =  $u_{n-1}$, and
$\overline{w}$ would have period $k$, which is impossible. $\Box$

\begin{pro} \label{aper}
If $s \in 2^{*}$ is an odd, aperiodic, maximal sequence, then the
subshift $(\Sigma_{{\cal D}^{\infty}(s)},\sigma)$ has not
chaotic limits.
\end{pro}
Proof: Let $U_{j}$ be sets from Lemma \ref{double}. Then
$\omega({\cal D}^{\infty}(s)) \subseteq U_{j}$, for $j>2$, so
$\omega({\cal D}^{\infty}(s))$ does not contain periodic points.
Suppose that $u \in \omega({\cal D}^{\infty}(s))$, $u \in A
\subseteq \Sigma_{{\cal D}^{\infty}(s)}$, and $\omega(A)$ is
chaotic. Then it contains a periodic point $v \in \omega(A)$.
Choose $j$ so that $v \not\in U_{j}$, and let $V$ be a
neighbourhood of $v$ disjoint from $U_{j}$. Then
$\sigma^{k}(U_{j}) \cap V = 0$ for all $k$, so $\omega(A)$ is not
transitive. $\Box$

\begin{thm} \label{chaot}
If $(I,g)$ is $S$-unimodal system, and ${\cal K}(g) = {\cal
D}^{\infty}(s)$ for some odd, aperiodic, maximal sequence $s$,
then $(I,g)$ has not chaotic limits.
\end{thm}
Proof: By Theorem \ref{nonper} there is a factorization
$\varphi: (\Sigma_{{\cal K}(g)}, \sigma) \rightarrow (I,g)$. Let
$y \in I$ be an eventually periodic point. Since ${\cal K}(g)$ is
aperiodic, $g^{i}(y) \neq c$ for all $i$, and therefore there
exists only one $u \in \Sigma_{{\cal K}(g)}$ with $y =
\varphi(u)$. Since $\omega({\cal K}(g))$ does not contain
periodic points by Proposition \ref{aper}, neither does
$\omega(c) =
\varphi(\omega({\cal K}(g))$. We prove that if $j>2$, then
$\varphi(U_{j})$ is a neighbourhood of $\omega(c)$. If $y \in
\omega(c)$, then there is unique $u \in \omega({\cal K}(g))$ with
$\varphi(u)=y$. There exists $k \leq 2^{j}ºsº$ such that
$\overline{0}_{ºk} \neq  u_{ºk} \neq 1\overline{0}_{ºk}$. Define
$u',u'' \in U_{j}$ by $u'_{ºk} = \overline{0}_{ºk}$, $u''_{ºk} =
1\overline{0}_{ºk}$, $u'_{i} = u''_{i} = u_{i}$ for $i \geq k$.
Then $u' \prec u \prec u''$ and $y=\varphi(u)$ is an inner point
of the interval $›\varphi(u'), \varphi(u'')! \subseteq
\varphi(U_{j})$. Since $\varphi(U_{j})$ is invariant, we get  the
result similarly as in Proposition \ref{aper}. $\Box$

\begin{cor}
Let $(I,g)$ be $S$-unimodal system, and ${\cal K}(g) = {\cal
D}^{\infty}(s)$ for some odd, aperiodic, maximal sequence $s$.
Then $(I,g)$ is not a factor of a finite automaton.
\end{cor}
Proof: Theorems \ref{limit} and \ref{chaot}.

\end{document}